\documentclass[aps,prc,twocolumn,floatfix,showpacs,a4paper,
nofootinbib,amsmath,amssymb]{revtex4-1}
\usepackage{graphicx}
\usepackage{dcolumn}
\usepackage{bm}
\usepackage{color}

\newcommand{\be}{\begin{equation}}
\newcommand{\ee}{\end{equation}}
\newcommand{\bea}{\begin{eqnarray}}
\newcommand{\eea}{\end{eqnarray}}
\newcommand{\bd}{\begin{displaymath}}
\newcommand{\ed}{\end{displaymath}}
\renewcommand{\vec}[1]{\mbox{\boldmath$#1$}}

\newcommand{\di}{{\rm d}}

\newcommand{\betav}{\boldsymbol{\beta}}

\newcommand{\Piv}{\boldsymbol{\Pi}}

\begin{document}
\title{Lambda Polarization in Peripheral Heavy Ion Collisions}

\author{F. Becattini$^{1,2}$, L.P.~Csernai$^3$, D.J. Wang$^{3,4}$}

\affiliation{
$^1$ University of Florence and INFN Florence, Italy\\
$^2$ Frankfurt Institute for Advanced Studies (FIAS), Johann Wolfgang Goethe University, Germany\\
$^3$ Institute of Physics and Technology, University of Bergen,
Allegaten 55, 5007 Bergen, Norway \\
$^4$ Key Laboratory of Quark and Lepton Physics (MOE) and Institute of Particle Physics,
Central China Normal University, Wuhan 430079, China.
}

\begin{abstract}
We predict the polarization of $\Lambda$ and $\bar\Lambda$ hyperons in peripheral 
heavy ion collisions at ultrarelativistic energy, based on the assumption of local 
thermodynamical equilibrium at freeze-out. The polarization vector is proportional to
the curl of the inverse temperature four-vector field and its length, of the order 
of percents, is maximal for particle with moderately high momentum lying on the 
reaction plane. A selective measurement of these particles could make $\Lambda$ 
polarization detectable. 
\end{abstract}

\date{\today}


\maketitle

\section{Introduction}

In peripheral high energy heavy ion collisions the system has a large angular momentum
\cite{Becattini}. It has been recently shown in hydrodynamical computation that 
this leads to a large shear and vorticity \cite{CMW13}. When the Quark-Gluon Plasma 
(QGP) is formed with low viscosity \cite{CKM}, interesting new phenomena may occur 
like rotation \cite{hydro1}, or even turbulence, in the form of a starting 
Kelvin-Helmholtz instability (KHI) \cite{hydro2,WNC13}, or other turbulent phenomena 
\cite{Stefan}. Furthermore, the large angular momentum may manifest itself in the 
polarization of secondary produced particles \cite{wang1,Becattini,torrieri}.
Recently, a formula for the polarization of weakly interacting particles with spin
$1/2$ at local thermodynamical equilibrium has been found in Ref.~\cite{BCZG} based on the 
extension of the Cooper-Frye formula to particles with spin. Provided that spin
degrees of freedom equilibrate locally, the polarization turns out to be proportional
to the vorticity of the inverse temperature four-vector field and can thus be
predicted in a full hydrodynamical calculation of the collision process ended by
the Cooper-Frye freeze-out prescription. 

Early measurements of the $\Lambda$ hyperon polarization \cite{Star07}, averaged
over a significantly large centrality range, indicated relatively small values, with 
an upper bound $| P_{\Lambda,\,\bar\Lambda} | \le 0.02$ averaging over all azimuthal 
angles of $\Lambda$ momentum. In this paper, we present a quantitative prediction 
of the $\Lambda,\bar \Lambda$ polarization, within a specific hydrodynamical calculation, 
at different centralities and its momentum dependence. 
At top RHIC energy ($\sqrt s _{NN}$ = 200 GeV), although the resulting polarization 
is of the order of 1-2\% on average, thus consistent with experimental bounds, it turns 
out to be the largest (around 7-9\%) for hyperons with moderately high momentum lying 
in the reaction plane. A selective measurement of $\Lambda$'s with few GeV momentum 
into the reaction plane could thus be able to show a finite polarization value, 
demonstrating that also spin degrees of freedom achieve local equilibrium and, in an 
indirect way, that vorticous flow is generated in peripheral heavy ion collisions.

The $\Lambda$ polarization arising from this pure thermo-mechanical effect (spin
degrees of freedom equilibration due to the equipartition principle, as mentioned
in the abstract) in principle competes with the polarization induced by the 
electromagnetic fields with the distinctive feature that the polarization 
vector induced by vorticity has the same orientation for particle and antiparticle, 
unlike for that induced by electromagnetic fields.
However, at the freeze-out stage, the magnetic field produced by the moving spectators
is estimated to be of the order of $2\,10^{10} T$ \cite{TonBra12} at $\sqrt s _{NN} = 200 
GeV$ so that the resulting polarization is of the order of $\mu_N g_\Lambda B/T \approx 10^{-6}$, 
i.e. at least four orders of magnitude less than our predicted value. 
The polarization of $\Lambda$ hyperons has been approached with different models
(e.g. \cite{wang1,barros}). Recently, Ref.~\cite{xnwang} has considered the local 
polarization of fermions in the plasma phase induced by the chiral anomaly, thus 
far with an unspecified transferring mechanism to final hadrons. We stress that in our 
approach the polarization of the observable hadrons is a consequence of the paradigm 
of local thermodynamical equilibrium; to be effective, the chiral anomaly should 
induce a modification of the velocity and temperature fields at the freeze-out.

\bigskip
 
\section{Polarization}

The $\Lambda$ polarization in the participant centre-of-mass frame, as a function 
of its momentum, reads (in units $c=K=1$) \cite{BCZG}:
\be\label{Pip}
  \Pi_\mu(p) = \hbar \epsilon_{\mu\rho\sigma\tau}\ \frac{p^\tau}{8 m}
  \frac{\int \di \Sigma_\lambda p^\lambda\, n_F( 1- n_F) \partial^\rho \beta^\sigma}
  {\int \di \Sigma_\lambda p^\lambda\, n_F} ,
\ee
where $\beta^\mu(x) = (1/T(x)) u^\mu(x)$ is the inverse temperature four-vector 
field, $n_F$ is the Fermi-J\"uttner distribution of the $\Lambda$, that is 
$1/(e^{\beta(x)\cdot{p}-\xi(x)}+1)$, being $\xi(x)=\mu(x)/T(x)$ with $\mu$ the
relevant $\Lambda$ chemical potential and $p$ its four-momentum. Because at the 
temperatures typical of freeze-out $\Lambda$ is quite dilute ($m_\Lambda \gg T$), 
the Pauli blocking factor, $(1{-}n_F)$, can be neglected in Eq. (\ref{Pip}). 
The very same formula, with the replacement $\xi \to -\xi$ applies to $\bar\Lambda$,
namely particles and antiparticles have the same polarization in the Boltzmann
approximation 
\footnote{Henceforth, unless otherwise stated, when referring to $\Lambda$
we mean both particle and antiparticle states.}.

\begin{figure}[ht]
\begin{center}
      \includegraphics[width=7.6cm]{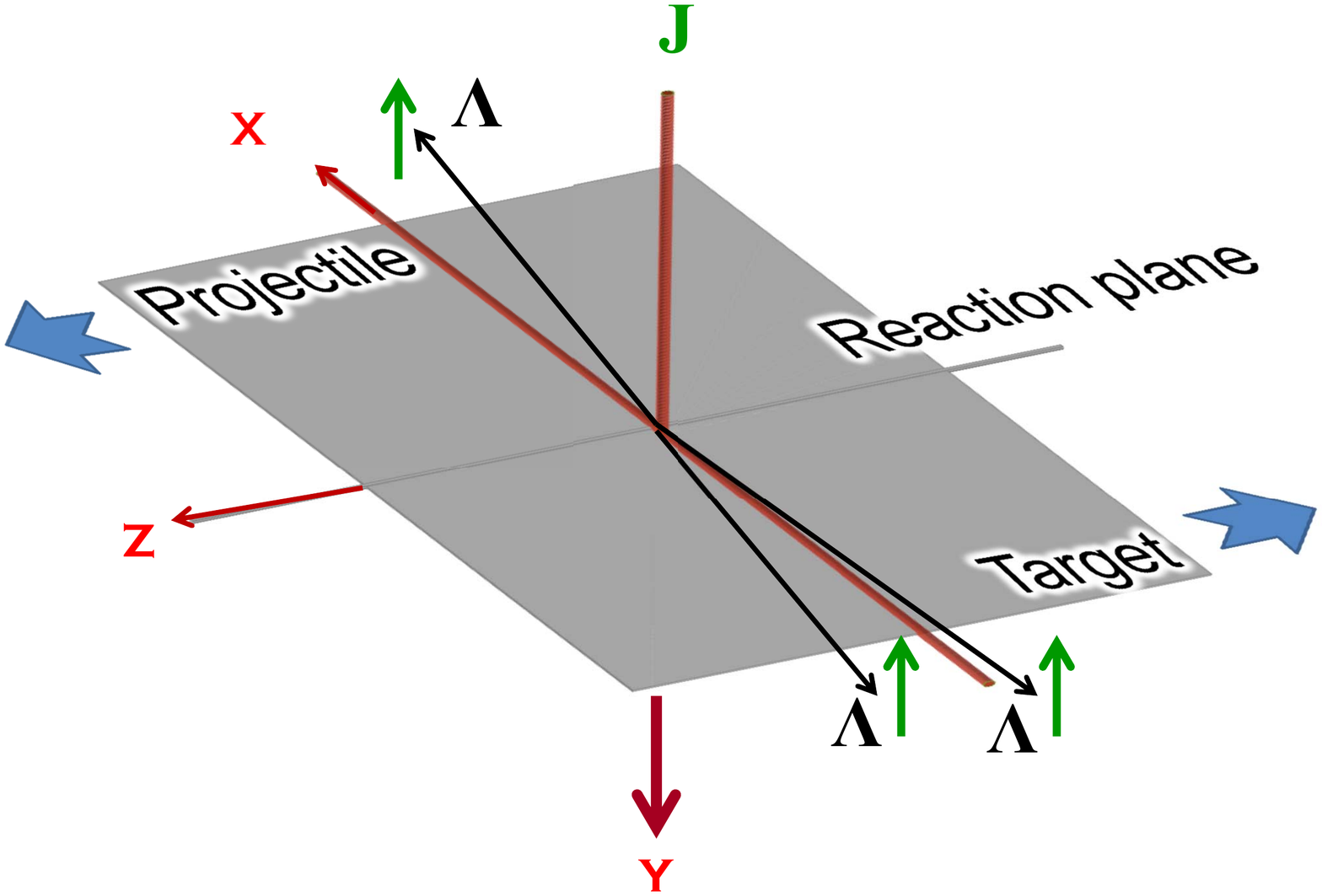}
\end{center}
\vspace{-0.5cm}
\caption{
(Color online)
Sketch of a peripheral heavy ion collisions at high energy. The $\Lambda$ 
polarization points essentially into the direction of the total angular
momentum ($-y$) of the interaction region, orthogonal to the reaction plane. 
$\Lambda$s with the largest polarization are emitted into the ($xz$) reaction plane.}
\vspace{-0.4cm}
\label{fig1}
\end{figure}

The polarization vector is then proportional to the antisymmetric part of the
gradient of the inverse temperature field, henceforth defined as {\em thermal vorticity}:
\be\label{thervor}
  \varpi^{\mu\nu} = \frac{1}{2} (\partial^\nu\beta^\mu-\partial^\mu\beta^\nu)
\ee
The spatial part of the polarization vector (\ref{Pip}) gives rise to three terms: 
\bea\label{Pipv}
 && \vec{\Pi}(p) = \frac{\hbar \varepsilon}{8m}
  \frac{\int \di \Sigma_\lambda p^\lambda \, n_F\ 
  (\nabla\times\vec{\beta})}{\int \di \Sigma_\lambda p^\lambda \,n_F} \nonumber \\
 && + \frac{\hbar {\bf p}}{8m} \times \frac{\int \di \Sigma_\lambda p^\lambda \,
 n_F\ (\partial_t \vec{\beta} + \nabla\beta^0)}
 {\int \di \Sigma_\lambda p^\lambda \,n_F} \,.
\eea
The last two terms on the right hand side, involving polar vectors, should vanish 
because of the overall parity invariance (achieved combining symmetry by reflection 
with respect to the reaction plane of the two colliding nuclei and invariance 
by rotation of $\pi$ around the axis orthogonal to the reaction plane). On the 
other hand, the first term, involving the spatial average of the curl of the $\betav$ 
field, which is an axial vector, is not ought to vanish; in fact it is a vector 
aligned with the total angular momentum direction, which is orthogonal to the 
reaction plane (see Fig.~\ref{fig1}). It should be pointed out that these formulae
apply to primary particles emitted from a locally equilibrated source. Secondary 
$\Lambda$s emitted from either strong or weak decays - most likely - will have 
a lower polarization inherited from their parent particles.

In the simplest scenario of an isochronous ($t=$const.) freeze-out at a given stage 
of the fluid dynamical expansion, according to Cooper-Frye prescription 
$\di \Sigma_\lambda p^\lambda\ \to \di V \varepsilon$, $\varepsilon = p^0$ 
being the $\Lambda$'s energy. In this case, the above formula simplifies to:
\be\label{Pipv2}
  \vec{\Pi}(p) = \frac{\hbar \varepsilon}{8m} \frac{\int \di V \, n_F\ 
  (\nabla\times\vec{\beta})}{\int \di V \,n_F}\ .
\ee  

The $\Lambda$ polarization is usually determined by measuring the angular distribution 
of the decay protons, which, in the $\Lambda$ rest frame is given by: 
$$
   \frac{1}{N}\frac{\di N}{\di \Omega^*} = \frac{1}{4\pi}
   \left( 1 + \alpha\, \Piv_0 \cdot \hat{\bf p}^* \right)
$$
where $\alpha = 0.647$, $\Piv_0$ is the polarization vector and $\hat{\bf p}^*$ is the 
direction of the decay proton, both in the $\Lambda$'s rest frame. The vector $\Piv_0$
can thus be obtained by Lorentz boosting to this frame the one in Eq.~(\ref{Pipv2}): 
\be
 \Piv_0(p) = \Piv(p) - \frac{{\bf p}}{\varepsilon(\varepsilon + m)} \Piv(p) \cdot {\bf p}
\label{LT}
\ee
where $(\varepsilon,{\bf p})$ is $\Lambda$ four-momentum and $m$ its mass. One
can readily realize that $\| \Piv_0 \| \le \| \Piv \|$ and equality is achieved
only if either ${\bf p} = 0$ (non-relativistic limit) or when ${\bf p} \cdot \Piv
= 0$. In both cases one has $\Piv_0(p) = \Piv (p)$. 
The above finding implies that maximal proper polarization of $\Lambda$ with finite 
momentum is achieved when they are {\em transversely} polarized. Thus, if $\Piv$ 
is directed along the total angular momentum ($-y$ direction in Fig.~\ref{fig1}),     
$\Lambda$'s having maximal polarization are those with momentum in the reaction 
plane or those with vanishing polar angle $\theta$ (normally undetectable) and, in 
this case, their proper polarization vector is aligned with the total angular momentum.   

\bigskip

\section{Hydrodynamical calculation}

The goal of the hydrodynamic calculation is to evaluate the thermal vorticity 
(\ref{thervor}) at the freeze-out. In this work we calculate it by using the Particle 
in Cell (PIC) fluid dynamic model, which provides us with the spacetime development 
of the flow of the QGP. The freeze-out is enforced by means of the Cooper-Frye 
prescription at a fixed laboratory time $t$, such that the average temperature
is $\approx 180$ MeV (see below). In comparison with Ref.~\cite{CMW13}, only the 
relativistic case is considered. 

For computational purposes, it is convenient to absorb the $\hbar$ constant into 
$\beta^{\mu}$ and redefine thermal vorticity as:
\be\label{thervor2}
  \varpi^{\mu \nu}=\frac{1}{2} (\partial^{\nu}\hat{\beta}^{\mu}-
  \partial^{\mu}\hat{\beta}^{\nu}),
\ee
where $\hat{\beta}^{\mu} \equiv \hbar \, \beta^{\mu}$. Thereby, $\varpi$ becomes
dimensionless. Note that in the thermal vorticity definition there is no projection 
of the derivatives transverse to the flow (the operator $\nabla_\mu = \partial_\mu
- u_\mu u_\nu \partial^\nu$), unlike in the usual definition of the vorticity
of the four-velocity field. 

\begin{figure}[ht] 
\begin{center}
      \includegraphics[width=7.6cm]{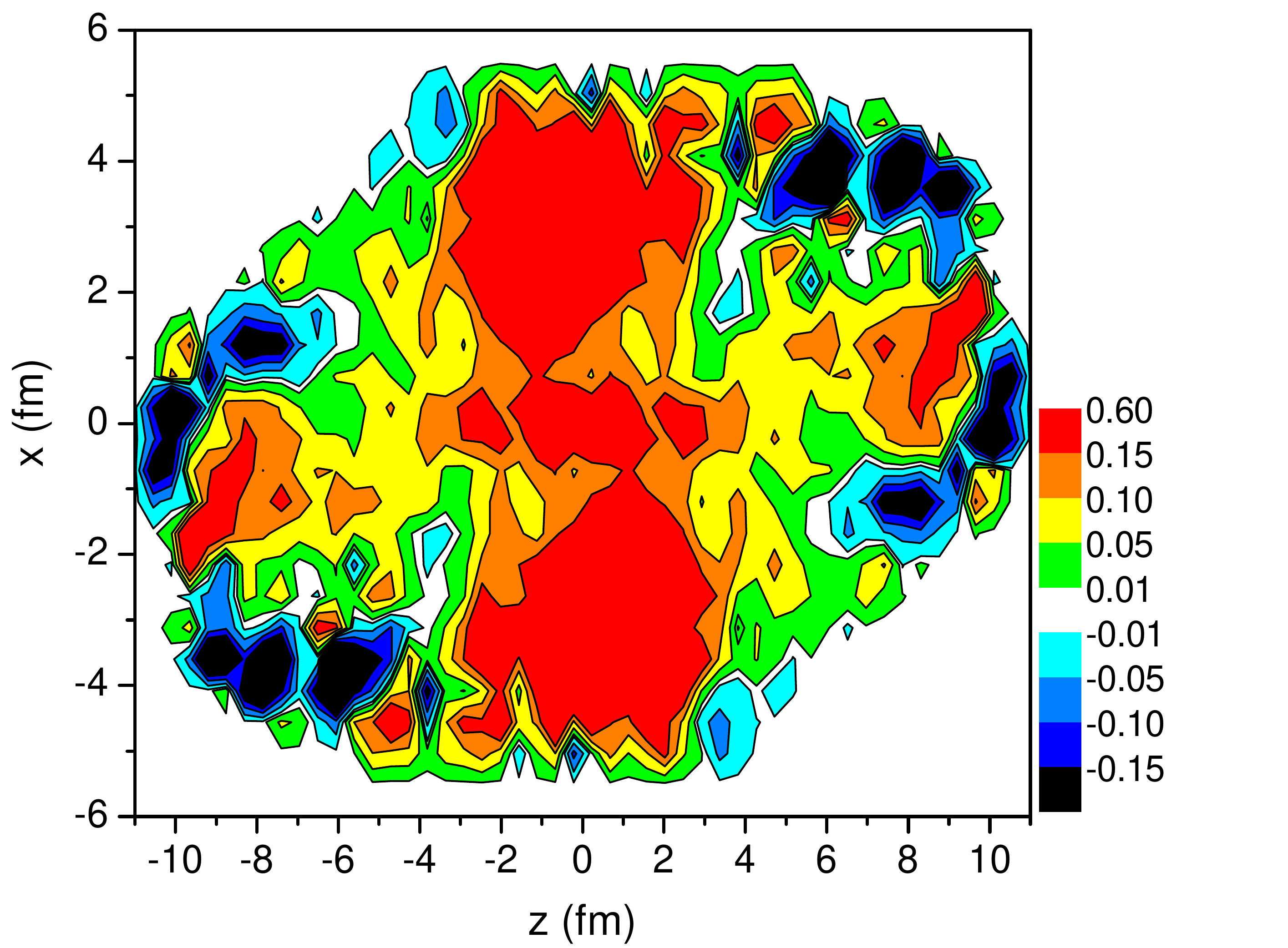}
\end{center}
\vspace{-0.3cm}
\caption{
(Color online)
The energy density weighted thermal vorticity, $\Omega_{zx}(x,z)$ 
of the inverse temperature four-vector field $\hat\beta^{\mu}$ (see text
for definition) calculated for all $[x-z]$ layers at t=4.75 fm/c,
correspoding to an energy density weighted temperature of 180 MeV.
The collision energy is $\sqrt{s_{NN}}=200$ GeV, $b=0.7\,b_{max}$. 
The cell size is $dx=dy=dz= 0.4375$ fm, while the average weighted vorticity is $\langle 
\Omega_{zx} \rangle = 0.0453 $.}
\vspace{-0.4cm}
\label{figt356}
\end{figure}

We present in Fig.~\ref{figt356} the $zx$ component of the thermal vorticity 
weighted with the energy density in the cell (that is $\Omega_{zx}({\rm cell}) 
= \varpi_{zx}({\rm cell}) \epsilon_{\rm cell} / \langle \epsilon \rangle$)
when the likewise weighted average temperature is 180 MeV, hence close to the 
freeze-out. The weighting with the energy density of the cell is described in 
detail in Ref.~\cite{CMW13}. 

From Fig.~\ref{figt356} it can be seen that, at the last time step presented, in 
the reaction plane we have already an extended area occupied by
matter. In case of peripheral reactions the multiplicity is relatively small, hence 
fluctuations in the reaction plane are considerable. In the relativistic case 
the outer edges show larger vorticity and random fluctuations are still strong.
The average vorticity is smaller for the smaller impact parameters and it has 
positive value in the center and negative value at the edges.

It should be pointed out that while the standard velocity field vorticity rapidly 
decreases with expansion \cite{CMW13}, thermal vorticity decrease is much slower
and at some peripheral points it even increases. This is due to the fact that the 
matter cools during the expansion, so the temperature in the denominator of $\beta^\mu$ 
decreases compensating for the decrease of velocity field vorticity with time.

One should also mention that, in our calculation, hydrodynamical evolution starts 
after a dynamical longitudinal expansion based on collective Yang-Mills dynamics. 
The initial longitudinal size of the system is about $2 \times 4$ fm, so the hydro
process starts  $\approx 4$ fm/c after the interpenetration of the two Lorentz 
contracted nuclei. Consequently the configuration in Fig. \ref{figt356} follows 
the interpenetration by about $8.5$ fm/c, which is the time at which the energy
density weighted average temperature is 180 MeV (see above).

\bigskip

\section{Results and Discussion}

The above described hydrodynamical calculation was performed for the conservative 
case (a) presented in Fig. \ref{figt356}, which 
represents an initial rotation without the enhancement due to KHI. To calculate
average polarization of $\Lambda$ hyperons, the thermal vorticity has been properly 
weighted with the Fermi-J\"uttner distribution $n_F$, according to Eq. (\ref{Pipv2}). 

The polarization vector, just as the flow vorticity, primarily points in the direction
of the total angular momentum ($-y$ in Fig.~\ref{fig1}). It depends on the $\Lambda$'s
momentum vector through the Fermi-J\"uttner distribution $n_F(p)$ (see Eq.~\ref{Pipv2}). 
It increases with $p_T$, and it is also sensitive to the flow properties and 
asymmetries; at $\pm p_y \approx 3$ GeV/c it is about $2$\%, while in the reaction 
plane at $p_x \approx 3$ GeV/c the polarization about $5$\%.
The proper polarization vector in the $\Lambda$ rest frame, determining the decay products 
angular distribution therein, is related to the polarization vector in the collision frame 
by Eq.~(\ref{LT}), which introduces a further dependence on the $\Lambda$'s direction,
as has been mentioned. Note that Eq.~(\ref{LT}) modifies the direction of $\Piv_0(p)$ 
with respect to $\Piv(p)$, except in the case when either ${\bf p}$ is lying on the 
reaction plane ($p_y=0$) or orthogonal to the reaction plane ($p_{x,z}=0$).

\begin{figure}[ht] 
\begin{center}
      \includegraphics[width=7.4cm]{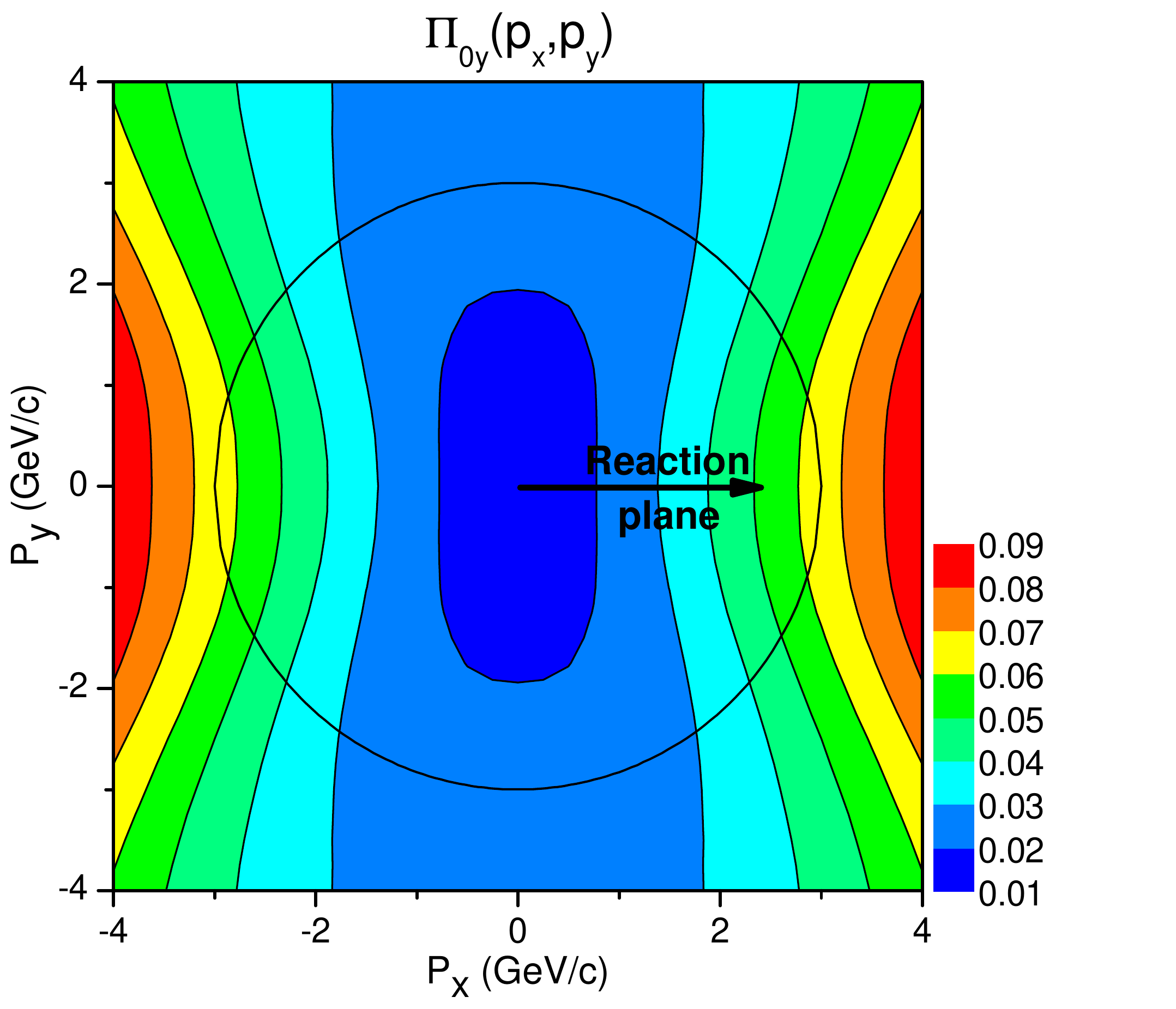}
      \includegraphics[width=7.4cm]{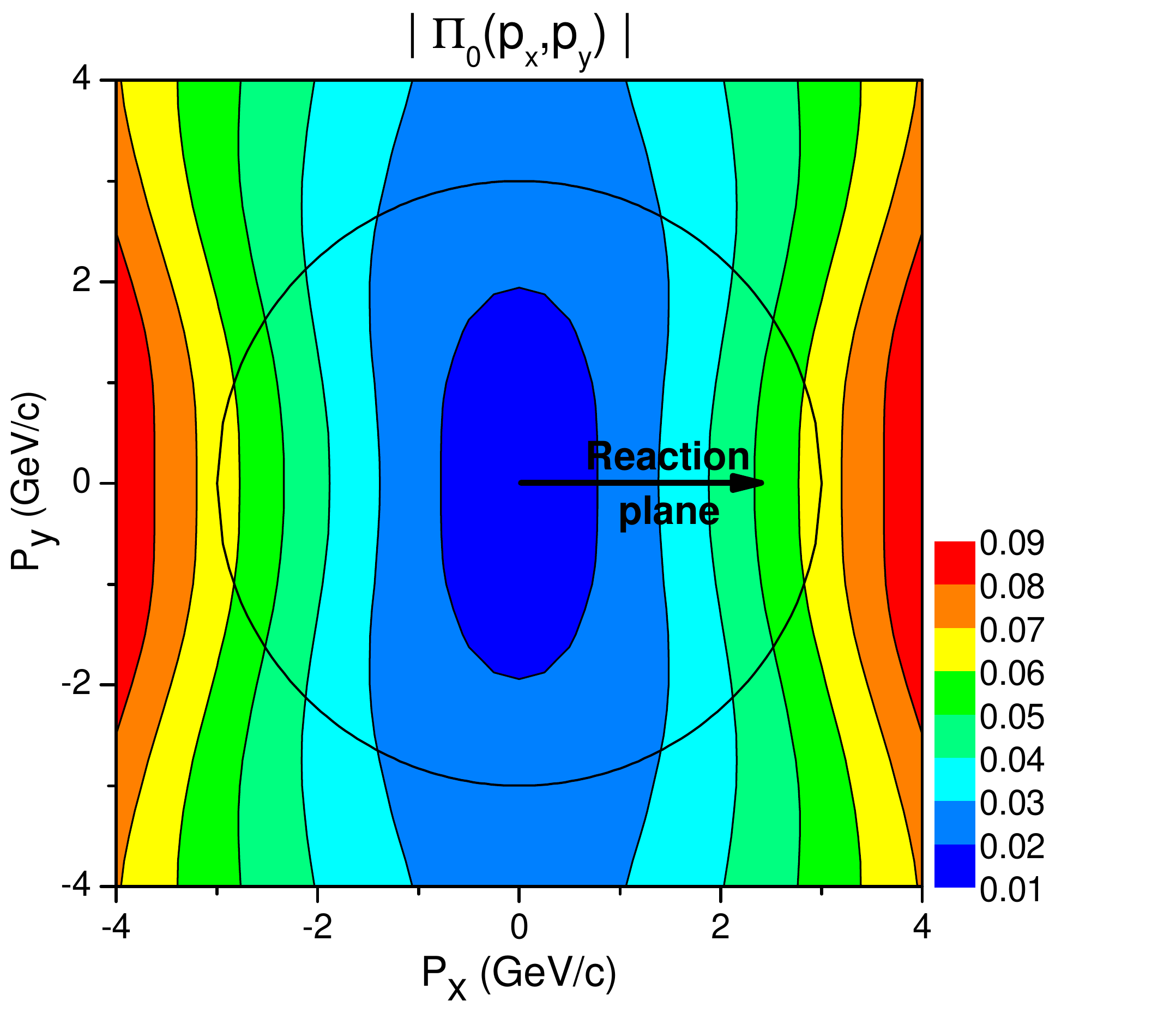}
\end{center}
\vspace{-0.5cm}
\caption{
(Color online) 
The dominant $y$-component and the modulus of the observable polarization, 
$\vec \Pi_0(\vec p)$ in the rest frame of the emitted $\Lambda$ as
a function of the $\Lambda$ momentum in the transverse plane at the
participant c.m. (i.e. at $p_z = 0$).}
\vspace{-0.4cm}
\label{figP0}
\end{figure}

The numerical results for the magnitude of the proper polarization and its projection
along the angular momentum axis $y$ are shown in Fig.~\ref{figP0}. It can be seen
that both increase as a function of transverse momentum and that their maximum values
are attained in the reaction plane ($x$ direction). The average value of polarization
is of the order of 1-2\% (consistent with RHIC bound), yet there are regions in momentum
space where polarization is significantly larger and could be found with a selective
measurement. The maximum up to $p_T > 3$ GeV/c is around 5\%, when $\Lambda$'s 
multiplicity is down by about 2 orders of magnitude compared to its top value, so 
that measuring polarization requires a sufficiently large statistics.
These results are significant even if our hydrodynamical model assumes the possible maximum
initial angular momentum; other (still realistic) models may have 10-20\% less,
yet without strongly reducing the final polarization value. It should also be noted
that the same hydro model shows the possible occurrence of the KHI, which enhance the 
effect by 10-20\% \cite{hydro2}.

It is important to stress that, in order to measure polarization, it is crucial
to determine the orientation of the reaction plane, that is of the total angular
momentum, on an event by event basis. As the polarization vector is oriented along
the total angular momentum, a misidentification of the orientation would average
to zero the measured polarization. A precise determination of the {\em direction} of the 
reaction plane is not as crucial because $\Lambda$ polarization does not vary much
in length and direction around it (being at a maximum, see Fig.~\ref{figP0}). In 
order to improve accuracy, the participant center of mass (c.m.) should also be determined, 
both in pseudorapidity and in the transverse plane. This is usually not easy due to 
the limited acceptance of the central 4$\pi$ detectors, but can be done by using 
the zero degree calorimeters with adequate correction factors as shown in 
Ref.~\cite{Eyyubova}, for the longitudinal c.m. The same can be done in the 
transverse direction too.

A possible background to the sought signal of ``hydrodynamical" polarization stems 
from polarized $\Lambda$'s emitted in single nucleon-nucleon (NN) collisions at the 
outer edge of the overlap region of the two colliding nuclei (the so-called {\em corona 
effect}). It must be first pointed out that in NN collisions only $\Lambda$'s are 
found to be polarized whereas $\bar\Lambda$'s have a polarization consistent with zero. 
Since our predicted polarization applies to both particle and antiparticle states, 
a non-vanishing $\bar\Lambda$ polarization would be free from this background. 
Nevertheless, we figure out that the NN background can be neglected also for $\Lambda$ 
particle.
Indeed, experimental observations show that $\Lambda$'s polarization scales with 
$x_F \equiv 2p/\sqrt{s}$ \cite{amsmith}, being $p$ its momentum in the NN 
centre-of-mass frame and that its magnitude strongly increases with $x_F$ 
\cite{lundberg}. At very low $x_F$, where our calculation is performed (with $y<1$ 
and $p_T$ up to 6 GeV, at the LHC energy scale of 1 TeV we have $x_F \simeq 0.07$)), 
the observed trend \cite{anselmino} indicates an approximate (generous) maximal 
polarization of 5\% for $p_T$ up to few GeVs. In order to estimate the impact of 
this background on the hydrodynamically originated polarization, one should
estimate the number of single NN collisions in the corona as a function of the 
number of participants nucleons $N_P$ in peripheral nuclear collisions. A calculation 
carried out by one of the authors \cite{becamann} with Glauber Monte-Carlo model 
at $\sqrt{s}_NN = 200$ GeV shows that for peripheral collisions with $N_P \simeq 100$
the number of nucleons undergoing single collisions in the corona is $N_{PC} \simeq 30$. 
According to STAR measurement \cite{starlambda}, for $N_P \simeq 100$, at midrapidity
the $\Lambda$ multiplicity is approximately $3.6 \times N_P$ times the one in pp
collisions at the same energy. Therefore, the fraction of $\Lambda$ coming from 
NN collisions with respect to the total production at $N_P=100$ can be estimated 
to be (see also eq.~(2) in ref.~\cite{becamann}) $(N_{PC}/2)/(3.6 N_P) = 15/360
\simeq 0.042$. This implies that at top RHIC energy, and even more so at LHC energy
where the fraction of corona collisions is lower, at most only about 4\% of 
$\Lambda$ hyperons come from NN collisions, and that their contribution to the 
measured polarization, at very low $x_F$, is at most $0.04 \times 0.05 = 0.002$, 
far below the signal level.

\section{Conclusions}

In conclusion, we have predicted the polarization of $\Lambda$ hyperons in relativistic 
heavy ion collisions at RHIC energy and its momentum dependence. Our calculation 
did not include the polarization of secondary $\Lambda$'s from decays of resonances 
or $\Xi$s which, most likely, will tend to dilute the signal. Still, the polarization
value may reach sizeable and detectable values of several percents for momenta
of some GeV's directed along the reaction plane. While the average value is predicted 
to be of the order of 1-2\%, in agreement with the experimental bound previously 
set at RHIC with about $10^7$ minimum bias Au-Au events \cite{Star07}, with the much larger 
statistics (at least a factor of 30) collected by RHIC in later runs \cite{run10}  the 
momentum differential measurement of $\Lambda$ and $\bar\Lambda$ polarization in
the direction along the reaction plane and at the participant c.m. should be feasible. 
We are also going to carry out similar calculations for the larger LHC energy.

The observation of a polarization arising from this thermo-mechanical effect of 
equipartition of angular momentum and in agreement with the predicted kinematic
features would be a striking confirmation of the achievement of local thermodynamical 
equilibrium (for the spin degrees of freedom too) of the matter created in relativistic 
heavy ion collisions. It would also indicate that significant vorticity and circulation
predicted in \cite{hydro1} may persist up to the freeze-out.

\bigskip

\section*{Acknowledgements}

This work was partly supported by Helmholtz International Center for FAIR. 
F.~Becattini, L.P.~Csernai and D.J.~Wang 
would like to acknowledge the kind hospitality at the 
Frankfurt Institute for Advanced Studies.


\end{document}